\documentclass[twocolumn,showpacs,amsmath,amssymb]{revtex4}
\usepackage{amssymb}
\usepackage[dvips]{graphicx}
\begin{document}

\title{High Temperature Superconductivity:  the explanation}
\author{A. S. Alexandrov}

\affiliation{Department of Physics, Loughborough University,
Loughborough LE11 3TU, United Kingdom\\}

\begin{abstract}
Soon after  the discovery of the first high temperature
superconductor by Georg Bednorz and Alex M\"uller in 1986 the late
Sir Nevill Mott  answering his own question   "Is there an
explanation?" [Nature \textbf{327} (1987) 185] expressed a view that
the Bose-Einstein condensation (BEC) of small bipolarons, predicted
by us in 1981, could be the one. Several authors then contemplated
BEC of real space tightly bound pairs, but with a purely electronic
mechanism of pairing rather than with the electron-phonon
interaction (EPI). However, a number of other researchers criticized
the bipolaron (or any real-space pairing) scenario as incompatible
with some angle-resolved photoemission spectra (ARPES), with
experimentally determined effective masses of carriers and
unconventional symmetry of the superconducting order parameter in
cuprates.  Since then the controversial issue of whether the
electron-phonon interaction (EPI) is crucial for high-temperature
superconductivity or weak and inessential has been one of the most
challenging problems of contemporary condensed matter physics.  Here
I outline some developments in the bipolaron theory suggesting that
the true origin of high-temperature superconductivity is found in a
proper combination of strong electron-electron correlations with a
significant finite-range (Fr\"ohlich) EPI, and that the theory is
fully compatible with the key experiments.

\end{abstract}


\maketitle \section{Real space and Cooper pairs}

There is still little consensus on the origin of high-temperature
superconductivity in cuprates \cite{muller} and other related
compounds. The only consensus there is is that  charge carriers are
bound into pairs with an integer spin.  Pairing of two fermionic
particles has been evidenced  in  cuprate superconductors
\cite{flux} from the quantization of magnetic flux in units of the
flux quantum $\phi_0 = h/2e$ .

A long time ago F. London suggested  that the remarkable superfluid properties of $%
^{4}$He were intimately linked to the Bose-Einstein ``condensation''
of the entire assembly of Bose particles \cite{lon0}. The crucial
demonstration that superfluidity was linked to the Bose particles
and the Bose-Einstein condensation came after experiments on liquid
$^{3}$He, whose atoms were fermions, which failed to show the
characteristic superfluid transition within a reasonable wide
temperature interval around the critical
temperature for the onset of superfluidity in $^{4}$He. In sharp contrast,$%
^{3}$He becomes a superfluid only below a very low temperature of some $%
0.0026$K. Here we have a superfluid formed from $pairs$ of two
$^{3}$He fermions below this temperature.

The three orders-of-magnitude difference between the critical
superfluidity temperatures of $^{4}$He and $^{3}$He kindles the view
that the Bose-Einstein condensation might represent the ``smoking
gun'' of high temperature superconductivity \cite{alebook}.
``Unfortunately'' electrons are fermions. Therefore, it is not
surprising at all that the first proposal for high temperature
superconductivity, made by Ogg Jr in 1946 \cite{ogg}, was the
pairing of individual electrons. If two electrons are chemically
coupled
together the resulting combination is a boson with the total spin $S=0$ or $%
S=1$. Thus an ensemble of such two-electron entities can, in
principle, be condensed into the Bose-Einstein superconducting
condensate. This idea was further developed as a natural explanation
of superconductivity by Schafroth \cite{scha}, and Butler and Blatt
in 1955 \cite{butl}.

However, with one or two exceptions \cite{edwards}, the
Ogg-Schafroth picture was condemned and practically forgotten
because it neither accounted quantitatively for the critical
parameters of  old (i.e. low $T_{c}$) superconductors, nor did it
explain the microscopic nature of the attractive force which could
overcome the natural Coulomb repulsion between two electrons which
constitute a Bose pair. The same model which yields a rather precise
estimate of the critical temperature of $^{4}$He leads to an utterly
unrealistic result for superconductors, $viz$, $T_{c}=10^{4}K$ with
the atomic density of electron pairs of about $10^{22}$ per
cm$^{3}$, and with the effective mass of each boson twice of the
electron mass, $m^{\ast \ast }=2m_{e}$.

The failure of this `bosonic' picture of individual electron pairs
became fully transparent when Bardeen, Cooper and Schrieffer (BCS)
\cite{bcs} proposed that two electrons in a superconductor were
indeed correlated in real space, but on a very large (practically
macroscopic) coherence length  of about $10^{4}$ times the average
inter-electron spacing. The BCS theory was derived from an early
demonstration by Fr\"{o}hlich \cite{fro} that conduction electrons
in states near the Fermi energy  attract each other on account of
their weak interaction with vibrating ions of a crystal lattice.
Cooper then showed that any two electrons were paired in the
momentum space  due to their quantum interaction (i.e the Pauli
exclusion principle) with all other electrons in the Fermi see.
These Cooper pairs  strongly overlap in the real space, in sharp
contrast with the model of non-overlapping real-space pairs
discussed earlier by Ogg and Schafroth, Butler and Blatt. Highly
successful for metals and alloys with a low  T$_{c}$, the BCS theory
led some theorists (e.g. Ref. \cite{and}) to the conclusion that
there should be no superconductivity above $30K$. While the
Ogg-Schafroth phenomenology predicted unrealistically high values of
$T_{c}$, the BCS theory left perhaps only a limited hope for the
discovery of new materials which could superconduct  at liquid
nitrogen or higher temperatures.

\section{Strong-coupling superconductivity beyond the BCS-Migdal-Eliashberg approximation}
It   became clear now, that the Ogg-Schafroth and BCS descriptions
are actually two opposite extremes of the electron-phonon
interaction. On a phenomenological level David Eagles \cite{eagles}
proposed pairing without superconductivity  in some temperature
range solving simultaneous equations for the BCS gap and for the
Fermi energy when an electron-electron attraction becomes greater
than some critical value, and that superconductivity sets in at a
lower temperature, of the order of the BEC temperature of the pairs
in  some low carrier-density compounds like SrTiO$_3$. Later on
extending the BCS theory towards the strong interaction between
electrons and ion vibrations, a Bose {\it liquid} of tightly bound
electron pairs surrounded by the lattice deformation (i.e. of {\it
small bipolarons}) was predicted \cite{aleran,aleran2}. Further
prediction was that high temperature superconductivity should exist
in the crossover region of the electron-phonon interaction strength
from the BCS-like to bipolaronic superconductivity \cite{ale0},
Fig.1.

\begin{figure}
\begin{center}
\includegraphics[angle=-90,width=0.55\textwidth]{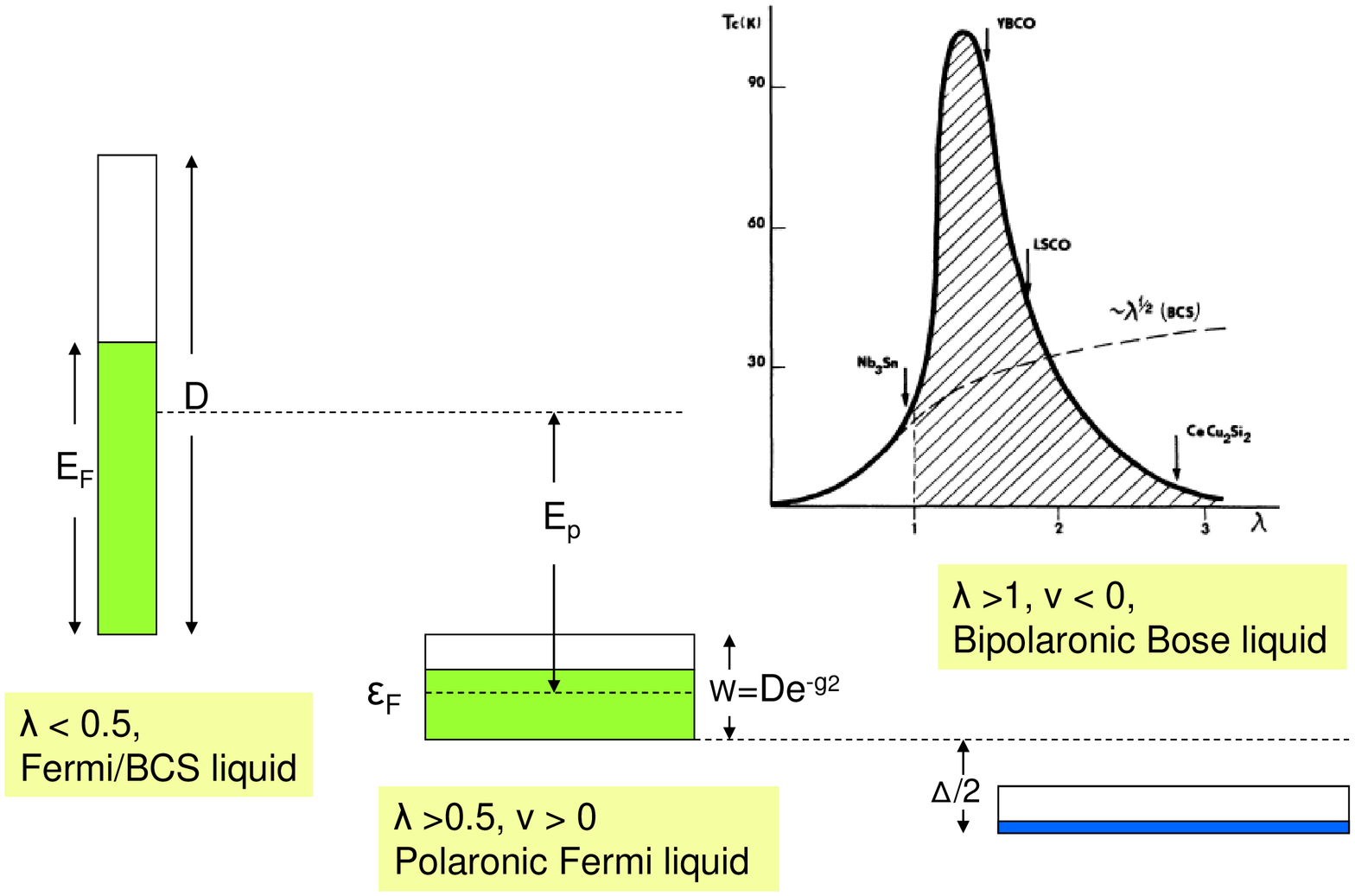}
\vskip -0.5mm \caption{(Color online) Breakdown of the BCS-ME
approximation due to the polaron bandwidth collapse at some critical
 coupling  $\lambda$ of the order of 1 (here v is the
polaron-polaron interaction). Inset shows the dependence of T$_c$
versus $\lambda$ in the polaron-bipolaron crossover region compared
with the BCS-ME dependence (dotted line). (Reproduced from
 Alexandrov A S 1988 New theory of strong-coupling superconductors and
high-temperature superconductivity of metallic oxides \emph{ Phys.
Rev. B} \textbf{38} 925, \copyright  American Physical Society,
1988.)}
\end{center}
\end{figure}

 Compared with the early Ogg-Schafroth view, two fermions (now small
polarons) are bound into a small bipolaron by the lattice
deformation. Such bipolaronic states,  at first sight,  have a mass
too large to be mobile. Actually earlier studies \cite{and0,chak}
considered small bipolarons as entirely localised objects. However
it has been shown
 analytically \cite{aleran,ale0,ale96,aubry} and using different
numerical techniques \cite{hague,bonca} that small bipolarons are
itinerant quasiparticles existing in the Bloch states at
temperatures below the characteristic phonon frequency which
coherently tunnel through the lattice with a reasonable effective
mass in particular, if the EPI is finite-range. As a result, the
superconducting critical temperature, proportional to the inverse
mass of a bipolaron, is reduced in comparison with the `ultra-hot'
local-pair Ogg-Schafroth superconductivity, but turns out  much
higher than the BCS prediction, Fig.1. A strong enhancement of T$_c$
in the crossover region from BCS-like polaronic to BEC-like
bipolaronic superconductivity is entirely due to a sharp increase of
the density of states in a narrow polaronic band \cite{ale0}, which
is missing in the so-called \emph{negative} Hubbard $U$ model
\cite{micnas,levin}. Quite remarkably Bednorz and M\"{u}ller noted
in their Nobel Prize lecture, that in their ground-breaking search
for High-T$_{c}$ superconductivity, they were stimulated and guided
by the polaron model. Their expectation
 was that if `{\it an electron and a surrounding lattice
distortion with a high effective mass can travel through the lattice
as a whole, and a strong electron-lattice coupling exists an
insulator could be turned into a high temperature superconductor}'
\cite{bed2}.

After we have shown  \cite{ale0}--- unexpectedly
 for many
 researchers--- that   the BCS-Migdal-Eliashberg (BCS-ME) theory  breaks down already at the EPI coupling $\lambda \gtrsim 0.5$ for any
 adiabatic ratio $\hbar \omega_0/E_F$, the multi-polaron physics has gained particular
attention \cite{aledev}.  The  parameter $\lambda
\hbar\omega_0/E_F$, which is supposed to be  small in the BCS-ME
theory \cite{mig,eli},  becomes in fact large at $\lambda \gtrsim
0.5$ since the electron bandwidth is narrowed and the Fermi energy,
$E_F$ is renormalised down exponentially below the characteristic
phonon energy, $\hbar \omega_0$, Fig.1 \cite{alebook}. Nevertheless,
as noted in the  unbiased comment by Jorge Hirsch \cite{hirsh}, in
order to explain the increasingly higher T$_c$s found in supposedly
'conventional' materials,  values of the electron-phonon coupling
constant $\lambda$ larger than $1$ have been used in the
conventional BCS-ME formalism.  This formalism completely ignores
the polaronic  collapse of the bandwidth, but regrettably continues
to be  used by some researchers irrespective of whether $\lambda$ is
small or large.

\section{Key pairing interaction and unconventional symmetry of the order parameter}
In general, the pairing mechanism of carriers
 could be not only "phononic" as in the BCS theory  or its
strong-coupling bipolaronic extension \cite{alebook}, but also
"excitonic", "plasmonic" ,
 "magnetic", "kinetic", or due to purely repulsive Coulomb
 interaction combined with an unconventional pairing symmetry of the order
 parameter \cite{alebook}.
Actually, following the original proposal by P. W. Anderson, many
authors \cite{tJ} assumed that the electron-electron interaction in
novel superconductors was very strong but repulsive and   provided
high $T_{c}$ without  phonons via e.g. superexchange,
spin-fluctuations, excitons or any other non-phononic mechanism. A
motivation for this concept can be found in the earlier work by Kohn
and Luttinger (KL) \cite{kohn}, who showed that the Cooper pairing
of fermions with a weak hard-core repulsion   was possible in a
finite orbital momentum state.  However the same work showed that
T$_{c}$ of hard-core repulsive fermions was  well below the mK
scale, and more importantly the KL  pairing with moderate values of
angular momenta ($p$ or $d$)  was impossible for \emph{charged}
fermions with the realistic finite-range Coulomb repulsion
\cite{lut,alegol} in disagreement  with some recent  claims
\cite{kiv}.  Also
 advanced  simulations with a
(projected) BCS-type trial wave function \cite{imada}, using the
 sign-problem-free Gaussian-Basis Monte Carlo algorithm
(GBMC)   showed  that the simplest repulsive  Hubbard model did not
account for high-temperature superconductivity  in the intermediate
and strong-coupling regimes either.

On the other hand some density functional (DFT) calculations
\cite{cohen,heid} found small EPI insufficient to explain high
critical temperatures within the BCS-ME framework, while  other
first-principles studies found large EPI in cuprates \cite{bauer}
and in recently discovered iron-based compounds \cite{yndurain}.  It
is a commonplace that DFT underestimates the role of the Coulomb
correlations and nonadiabatic effects, predicting an anisotropy of
electron-response functions much smaller than that experimentally
observed in the layered high-$T_{c}$ superconductors. Adiabatic DFT
calculations  could not explain the optical infrared c-axis spectra
and the corresponding electron-phonon coupling in the metallic state
of the cuprates.  On the other hand, these spectra are well
described within the nonadiabatic response approach of
Ref.\cite{bauer}. There is a strong nonlocal polar EPI along the
c-axis in the cuprates together with optical conductivity as in an
ionic insulator even in the well-doped ""metallic" state
\cite{bauer}. The inclusion of a short-range repulsion (Hubbard $U$)
via the LDA+U algorithm \cite{zhang} also significantly enhances the
EPI strength due to a poor screening of some particular phonons.
Substantial isotope effects on the carrier mass and a number of
other independent observations (see e.g.\cite{asazhao} and
references therein) unambiguously show that lattice vibrations play
a significant although unconventional role in high-temperature
superconductors. Overall, it seems plausible that the true origin of
high-temperature superconductivity should be found in a proper
combination of strong electron-electron correlations with a
significant EPI \cite{acmp}.

 We have recently calculated the EPI strength, the
phonon-induced electron-electron attraction, and the carrier mass
renormalization in layered superconductors at different doping using
a continuum approximation for the renormalized carrier energy
spectrum and the RPA dielectric response function \cite{alebra2010}.

If, for instance we start with a parent insulator as
La$_{2}$CuO$_{4}$,  the magnitude
of the Fr\"{o}hlich EPI is unambiguously estimated using the static, $%
\epsilon _{s}$ and high-frequency, $\epsilon _{\infty }$ dielectric
constants \cite{ale96,alebra}. To assess its strength, one can apply
an expression for the polaron binding energy (polaronic level shift)
$E_{p}$,
which depends only on the measured $\epsilon _{s}$ and $\epsilon _{\infty }$,%
\begin{equation}
E_{p}={\frac{e^{2}}{{2\epsilon _{0}\kappa
}}}\int_{BZ}{\frac{d^{3}q}{{(2\pi )^{3}q^{2}}}}.  \label{shift}
\end{equation}%
Here, the integration goes over the Brillouin zone (BZ), $\epsilon
_{0}\approx 8.85\times 10^{-12}$ F/m is the vacuum permittivity, and
$\kappa =\epsilon _{s}\epsilon _{\infty }/(\epsilon _{s}-\epsilon
_{\infty })$. In the parent insulator, the Fr\"{o}hlich interaction
alone provides the binding energy of two holes, $2E_{p}$, an order
of magnitude larger than any magnetic interaction ($E_{p}=0.647$ eV
in La$_{2}$CuO$_{4}$ \cite{alebra}). Actually, Eq.(\ref{shift})
underestimates the polaron binding energy, since
the deformation potential and/or molecular-type (e.g. Jahn-Teller \cite{mul}%
) EPIs are not included.

It was argued earlier \cite{ale96} that the interaction with c-axis
polarized phonons in cuprates would remain strong also at finite
doping due to a poor screening of high-frequency electric forces as
confirmed in some pump-probe \cite{boz2,dragan} and photoemission
\cite{shen,meevasana} experiments. However, a quantitative analysis
of the doping dependent EPI remained elusive because the dynamic
dielectric response function, $\epsilon (\omega ,\mathbf{q})$ was
 unknown. Recent observations of the quantum magnetic
oscillations in some underdoped \cite{und} and overdoped \cite{over}
cuprate superconductors are opening up a possibility for a
quantitative assessment of EPI in these and related doped ionic
lattices with the quasi two-dimensional (2D) carrier energy
spectrum. The oscillations revealed cylindrical Fermi surfaces,
enhanced effective masses of carriers (ranging from $2m_{e}$ to
$6m_{e}$) and the astonishingly low Fermi energy, $E_F$, which
appears to be well below 40 meV in underdoped Y-Ba-Cu-O \cite{und}
and less or about 400 meV in heavily overdoped Tl2201 \cite{over}.
Photoemission spectroscopies (\cite{shen,meevasana} and references
therein) do not show small Fermi-surface pockets and there are
alternative interpretations of slow magnetic oscillations in
underdoped cuprates unrelated to Landau quantization, for example
\cite{aleosc}. However, a poorly screened strong EPI
\cite{alebra2010}
 is not sensitive to particular band structures and Fermi
surfaces, but originates in the low Fermi energy, which is supported
by other independent experiments \cite{aleadi}. Such low Fermi
energies  make the Migdal-Eliashberg (ME) adiabatic approach to EPI
 inapplicable in
these compounds. Indeed, the ME non-crossing approximation breaks down at $%
\lambda \hbar \omega _{0}/E_{F}>1$ when the crossing diagrams become
important. The characteristic oxygen vibration energy is about
$\hbar \omega _{0}=80$ meV in oxides, therefore the ME theory cannot
be applied even for a weak EPI with the coupling constant $\lambda
<0.5$. In
the strong coupling regime, $\lambda \gtrsim 0.5,$ the effective parameter $%
\lambda \hbar \omega _{0}/E_{F}$ becomes large irrespective of the
adiabatic ratio, $\hbar \omega _{0}/E_{F}$, because the Fermi energy
shrinks exponentially due to the polaron narrowing of the band
\cite{ale0}. Since carriers in cuprates are in the non-adiabatic
(underdoped) or near-adiabatic (overdoped) regimes, $E_{F}\lesssim
\hbar \omega _{0}$, their energy spectrum renormalized by EPI and
the polaron-polaron interactions can be found with the familiar
small-polaron canonical transformation at \emph{any coupling
}$\lambda $ \cite{alekor}.

With doping the attraction and the polaron mass drop
\cite{alebra2010}. Nevertheless,  on-site  and  inter-site
attractions induced by EPI remain well above the superexchange
(magnetic) interaction $J$ (about 100 meV) at any doping since the
non-adiabatic carriers cannot fully screen high-frequency electric
fields.  The polaron mass
\cite{alebra2010} agrees quite well with the experimental masses \cite%
{und,over}. Decreasing the phonon frequency lowers the attraction
and increases the polaron mass in underdoped compounds with little
effect on both quantities at overdoping. Hence the Fr\"{o}%
hlich EPI with high-frequency optical phonons turns out to be the
key pairing interaction in underdoped cuprates and remains the
essential player at overdoping. What is more surprising is that EPI
is clearly beyond the BCS-ME approximation since its magnitude is
larger than or comparable with the Fermi energy and the carriers are
in the non-adiabatic or near-adiabatic regimes.

Together with the deformation potential and Jahn-Teller EPIs, the Fr\"{o}%
hlich EPI overcomes the direct Coulomb repulsion at distances
compared with the lattice constant even without any retardation
\cite{ale0}. Since EPI is not local in the nonadibatic electron
system with poor screening it can provide the d-wave symmetry of the
pairing state \cite{aledwave}.  Remarkably  the internal symmetry of
an individual bipolaron in underdoped regime could be different from
the symmetry of the Bose-Einstein condensate if the pairing takes
place with nonzero center-of-mass momentum \cite{alesym}. All these
conditions point to a crossover from the bipolaronic to polaronic
superconductivity \cite{ale0} in  cuprates with doping.

\section{Pseudogap, superconducting gap,  ARPES and tunnelling spectra of cuprate superconductors}
A detailed microscopic theory capable of describing unusual ARPES
and tunnelling data has so far remained elusive. Soon after  the
discovery of high-T$_c$ superconductivity \cite{muller}, a number of
tunnelling, photoemission, optical, nuclear spin relaxation, and
electron-energy-loss spectroscopies discovered an anomalously large
gap in cuprate superconductors existing well above the
superconducting critical temperature, T$_c$. The gap, now known as
the pseudogap, was originally assigned \cite{aleray} to the binding
energy of real-space preformed hole pairs - small bipolarons - bound
by a strong EPI.  Since then, many alternative explanations of the
pseudogap have been proposed.

The present-day scanning tunnelling  (STS)
\cite{gomes,davis2009,kato}, intrinsic tunnelling \cite{krasnov} and
angle-resolved photoemission (ARPES) \cite{shen2007,shen2009}
spectroscopies have offered a tremendous advance into the
understanding of the pseudogap phenomenon in cuprates and some
related compounds. Both extrinsic (see \cite{gomes,kato} and
references therein) and intrinsic \cite{krasnov} tunnelling as well
as high-resolution ARPES \cite{shen2007} have  found another energy
scale, reminiscent of a BCS-like \textquotedblleft superconducting"
gap that opens at T$_c$ accompanied by the appearance of
Bogoliubov-like quasi-particles \cite{shen2007} around the node.
Earlier experiments with a time-resolved pump-probe demonstrated two
distinct gaps, one  a temperature independent pseudogap and the
other a BCS-like gap \cite{dem}.  Another remarkable observation is
the spatial nanoscale inhomogeneity of the pseudogap observed with
STS \cite{gomes,davis2009,kato} and presumably related to an
unavoidable disorder in doped cuprates. Essentially, the doping and
magnetic field dependence of the superconducting gap compared with
the pseudogap and their different real space  profiles have prompted
an opinion that the pseudogap is detrimental to superconductivity
\cite{krasnov}.

Without a detailed microscopic theory which could describe highly
unusual tunnelling and ARPES spectra, the relationship between the
pseudogap and the superconducting gap has remained a mystery
\cite{shen2007}. Recently  we have developed the bipolaron theory of
ARPES \cite{alekim} and   tunnelling \cite{alejoannePRL,
alejoannePRB} by taking into account real-space pairing, coherence
effects in a single-particle excitation spectrum and disorder. Our
theory accounts for major peculiarities in extrinsic and intrinsic
tunnelling in  cuprate superconductors, and in ARPES.

 Real-space pairs, whatever
the pairing interaction is, can be described as a charged Bose
liquid on a lattice, if the carrier density is relatively small
avoiding their overlap \cite{alebook}. The superfluid state of such
a liquid is the true Bose-Einstein condensate (BEC), rather than a
coherent state of overlapping Cooper pairs. Single-particle
excitations of the liquid are thermally excited single polarons
propagating in a doped insulator band or are localised by
impurities. Different from the BCS case, their {\it negative}
chemical potential, $\mu$, is found outside the band by about half
of the bipolaron binding energy, $\Delta _{p}$, both in the
superconducting and normal states \cite{alebook}. Here, in the
superconducting state (T$<$T$_{c}$), following Ref.\cite{aleand} one
takes into account that polarons interact with the condensate via
the same potential that binds the carriers.  As in the BCS case the
single quasi-particle energy spectrum, $\epsilon_{\nu}$, is found
using the Bogoliubov transformation, $\epsilon_{\nu}=\left[
\xi_{\nu}^2+\Delta _{c \nu}^{2}\right] ^{1/2}$. However,  this
spectrum is different from the BCS quasi-particles because the
chemical potential is negative with respect to the bottom of the
single-particle band, $\mu =-\Delta _{p}$. A single-particle gap,
$\Delta$, is defined as the minimum of $\epsilon_{\nu}$.  Without
disorder, for a point-like pairing potential with the s-wave
coherent gap, $\Delta_{c {\bf k}}\approx \Delta_{c}$, one has
\cite{aleand} $\Delta(T)=\left[ \Delta _{p}^{2}+\Delta
_{c}(T)^{2}\right] ^{1/2}$. The full gap varies with temperature
from $\Delta (0)=\left[ \Delta_{p}^{2}+\Delta _{c}(0)^{2}\right]
^{1/2}$ at zero temperature down to the temperature independent
$\Delta=\Delta _{p}$ above T$_{c}$, which qualitatively describes
some earlier  and more recent \cite{krasnov} observations including
the Andreev reflection in cuprates (see \cite{aleand} and references
therein).

\begin{figure}
\begin{center}
\includegraphics[angle=-90,width=0.50\textwidth]{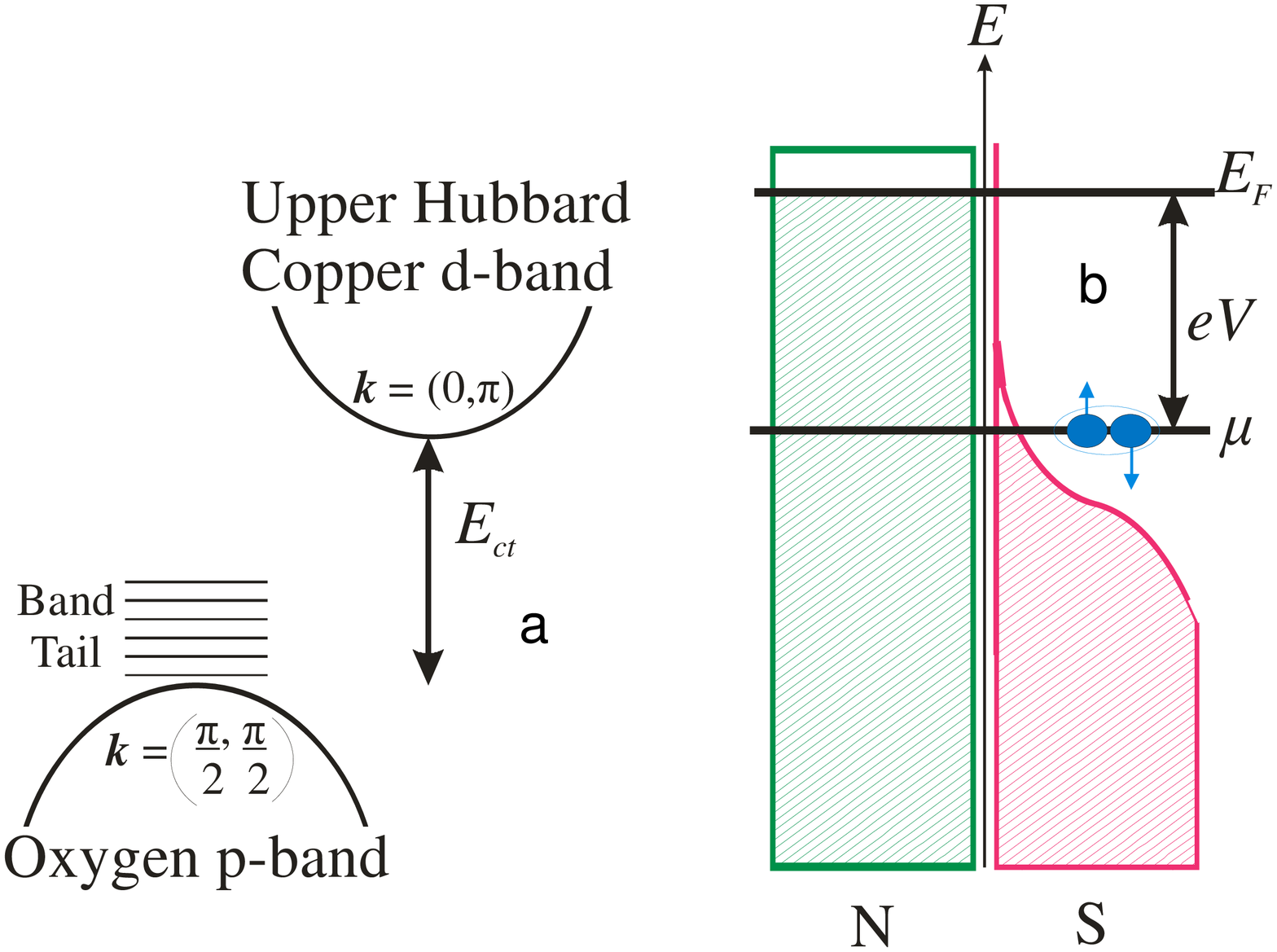}
\vskip -0.5mm \caption{(Color online) LDA+GTB energy band structure
of the cuprates with the impurity localised states \cite{alekim}
shown as horizontal lines in ($a$); NS model densities of states
$(b)$ showing  the bandtail  in the bosonic superconductor.
(Reproduced from Alexandrov A S and J. Beanland J 2010
Superconducting Gap, Normal State Pseudogap, and Tunneling Spectra
of Bosonic and Cuprate Superconductors \emph{Phys. Rev. Lett.}
\textbf{104} 026401 , \copyright  American Physical Society, 2010.)}
\end{center}
\end{figure}

To calculate ARPES and the tunnelling conductance  we  adopted the
first-principle \textquotedblleft LDA+GTB" band structure \cite{kor}
amended with impurity band-tails \cite{alekim}, Fig.2. It explains
the charge-transfer gap, $E_{ct}$, sharp \textquotedblleft
quasi-particle" peaks near $(\pi/2,\pi/2)$ of the Brillouin zone and
a high-energy \textquotedblleft waterfall" observed by ARPES in
underdoped cuprate superconductors \cite{alekim}. The chemical
potential is found in the single-particle bandtail within the
charge-transfer gap at the bipolaron mobility edge, Fig.2b, in
agreement with the S-N-S tunnelling experiments \cite{boz}. Such a
band structure explains an insulating-like low temperature
normal-state resistivity  as well as many  other unusual
normal-state properties of underdoped cuprates \cite{alebook}.

The bipolaron theory captures key unusual signatures of the
experimental tunnelling conductance in  cuprates, such as the
low-energy coherent gap, the high-energy pseudogap, and the
asymmetry \cite{alejoannePRL}. In the case of atomically resolved
STS one should use a $local$ bandtail DOS $\rho(E, {\bf r})$,  which
depends on different points of the scan area ${\bf r}$ due to a
nonuniform dopant distribution, rather than an averaged DOS or
spectral functions measured with ARPES. As a result the pseudogap
shows nanoscale inhomogeneity, while the low-energy coherent gap is
spatially uniform, as observed \cite{kato}. Increasing doping level
tends to diminish the bipolaron binding energy, $\Delta_p$,  since
the pairing potential becomes weaker due to  a partial screening of
EPI with low-frequency phonons \cite{alekabmot,alebra2010}. However,
the coherent gap, $\Delta_c$, which is the product of the pairing
potential and the square root of the carrier density \cite{aleand},
can remain about a constant or even increase with doping, as also
observed \cite{kato}.

\section{Concluding remarks on lattice (bi)polarons in high-temperature
superconductors}
 A  growing number of  observations tell us that
high-$T_{c}$ cuprate superconductors \cite{muller} are not the
conventional BCS superconductors \cite{bcs}, but represent a
realisation of strong-coupling non-adiabatic polaronic and
bipolaronic superconductivity \cite{alebook}. The fundamental origin
of such strong departure of superconducting cuprates from
conventional BCS behaviour stems from the poorly screened Fr\"ohlich
EPI
 of the order of 1 eV, routinely neglected in the Hubbard $U$ and $t-J$ models \cite{tJ}.
This interaction with
 optical phonons is poorly screened  because the charge
 carriers are found in the non-adiabatic or near adiabatic regime with
 their plasmon frequency  below or about the characteristic
frequency of optical phonons.  Since
 screening is poor, the magnetic interaction remains
small compared with the Fr\"ohlich EPI at any doping of cuprates.
Consequently, in order to generate an adequate theory of
high-temperature superconductivity,  finite-range Coulomb repulsion
and the Fr\"ohlich EPI must be treated on an equal footing. When
both interactions are strong compared with the kinetic energy of
carriers, our theory  predicts the low-energy state in the form of
mobile inter-site bipolarons at underdoping and mobile small
polarons at overdoping \cite{alebook}.

There is abundant independent evidence in favor of (bi)polarons
\cite{aledev} and 3D BEC in cuprate superconductors
\cite{alenev,alerev}. The  substantial isotope effect on the carrier
mass \cite{zhao,zhao2,khas,annet} predicted for (bi)polaronic
conductors in \cite{aleiso} is perhaps the most compelling evidence
for (bi)polaronic carries in cuprate superconductors. High
resolution ARPES \cite{lanzara,shen,meevasana} provides another
piece of evidence for the strong EPI in cuprates and related doped
layered compounds apparently with c-axis-polarised optical phonons.
These as well as
  tunnelling \cite{zhaot,bozt}
spectroscopies of cuprates, and recent pump-probe experiments
\cite{boz2,dragan}  unambiguously show that the Fr\"{o}hlich EPI is
important in those highly polarizable ionic lattices.

A parameter-free estimate of the Fermi energy using the
magnetic-field penetration depth \cite{aleadi} and the magnetic
quantum oscillations \cite{und} found its very low value, $\epsilon
_{F}\lesssim 50$ meV  supporting the real-space  pairing in
underdoped cuprate superconductors.

Magnetotransport  and thermal magnetotransport
 data strongly support preformed bosons in cuprates.
In particular,
 many high-magnetic-field studies revealed a non-BCS upward
curvature of the upper critical field $H_{c2}(T)$ (see \cite{ZAV}
for a review of experimental data),  predicted for the Bose-Einstein
condensation of charged bosons in the magnetic field \cite{aleH}.
The Lorenz
number, $L= \kappa _{e}/T\sigma $ differs significantly from the Sommerfeld value $%
L_{e}$ of the standard Fermi-liquid theory, if carriers are
double-charged bosons \cite{alemott}. Here $\kappa _{e}$, and
$\sigma $  are  electron thermal and electrical conductivities,
respectively. Ref.\cite{alemott} predicted a rather low Lorenz
number for bipolarons, $L\approx 0.15L_{e}$, due to the double
elementary charge of bipolarons, and also due to their nearly
classical distribution function above $T_{c}$. Direct measurements
of the Lorenz number  using the thermal Hall effect \cite{zha}
produced the value of $L$ just above $T_{c}$ about the same as
predicted by the bipolaron model.

Single polarons, \emph{localised }within an impurity band-tail,
coexist with bipolarons in the charge-transfer doped Mott-Hubbard
insulator. They account for  sharp \textquotedblleft quasi-particle"
peaks near $(\pi/2,\pi/2)$ of the Brillouin zone and high-energy
\textquotedblleft waterfall" effects observed with ARPES in cuprate
superconductors \cite{alekim}. This band-tail model also accounts
for two energy scales in  ARPES and in the extrinsic and intrinsic
tunnelling, their temperature and doping dependence, and for the
asymmetry and inhomogeneity of extrinsic tunnelling spectra of
cuprates \cite{alejoannePRL}. On the other hand, essentially
different doping and magnetic field dependence of the
superconducting gap compared with the pseudogap and their different
real-space profiles have prompted the  opinion \cite{krasnov} that
the pseudogap is not connected to the so-called "preformed" Cooper
pairs advocated by Emery and Kivelson \cite{emery} as an alternative
to the BEC of real-space pairs. The unusual normal state
diamagnetism uncovered by torque magnetometery
 has also been convincingly explained as the
normal state (Landau) diamagnetism of charged bosons \cite{aledia}.

Overall the real-space pairing seems to be a remarkable feature of
cuprates no matter what the microscopic pairing mechanism is. A
lattice disorder introduces additional complexity to the problem
since an  interference of impurity potential with the lattice
distortion, which accompanies the polaron movement,  contributes to
the polaron and bipolaron localization \cite{phillips2003}.
 Self-organized discrete
dopant networks \cite{phillips}  lead to multiscale complexity for
key materials as well. However the detailed microscopic physics of
the bosonic many body state seems to be irrelevant for fitting their
electrodynamic properties \cite{aledia2}.

As was emphasized in a number of early qualitative \cite{alenev}
(1990s) and more recent numerical studies \cite{trugman,mis} of
strongly correlated electrons with a significant EPI, the
anti-ferromagnetic spin fluctuations facilitate a doping-induced
lattice polaron formation profoundly lowering the required strength
of the bare EPI, but playing virtually no role in pairing compared
with the  EPI \cite{alebra2010,vidmar}. It is quite surprising that
despite clear evidence for lattice polarons in cuprate and related
superconductors, there are yet opinions (e.g. \cite{tJ}) suggesting
that EPI is inessential or that polaron formation does not help, but
hinder the pairing instability, very much in contrast to the notion
advanced by this and other authors that in fact lattice polarons
explain high-T$_c$ superconductivity.

\end{document}